\documentclass[twocolumn,aps,showpacs,preprintnumbers,amsmath,amssymb]{revtex4}

\usepackage{graphicx}
\usepackage{dcolumn}
\usepackage{bm}

\begin{document}

\title{Weak ferromagnetism and other instabilities of the
two-dimensional $t$-$t^{\prime }$ Hubbard model at Van Hove fillings}
\author{V.~Hankevych$^{1,2}$} 
\author{B.~Kyung$^1$}
\author{A.-M.~S.~Tremblay$^{1,3}$} 
\affiliation{$^{1}$D\'{e}partement
de physique and Regroupement qu\'{e}b\'{e}cois sur les mat\'{e}riaux
de pointe, Universit\'{e} de Sherbrooke, Sherbrooke Qu\'{e}bec J1K
2R1, Canada\\ 
$^2$Department of Physics, Ternopil State Technical
University, 56 Rus'ka St., UA-46001 Ternopil, Ukraine\\ 
$^3$Institut
canadien de recherches avanc\'{e}es, Universit\'{e} de Sherbrooke,
Sherbrooke Qu\'{e}bec, J1K 2R1, Canada} 
\date{\today }

\begin{abstract}
We investigate magnetic and superconducting instabilities of the
two-dimensional $t$-$t^{\prime }$ Hubbard model on a square lattice at
Van Hove densities from weak to intermediate coupling by means of the
Two Particle Self-Consistent approach (TPSC). We find that as the
next-nearest-neighbor hopping $|t^{\prime }|$ increases from zero, the
leading instability is towards an incommensurate spin-density wave
whose wave vector moves slowly away from $\left( \pi ,\pi
\right)$. For intermediate values of $|t^{\prime }|$, the leading
instability is towards $ d_{x^{2}-y^{2}}$-wave superconductivity. For
larger $|t^{\prime }|>0.33t$, there are signs of a crossover to
ferromagnetism at extremely low temperatures. The suppression of the
crossover temperature is driven by Kanamori screening that strongly
renormalizes the effective interaction and also causes the crossover
temperature to depend only weakly on $t^{\prime}$. Electronic
self-energy effects for large $|t^{\prime }|$ lead to considerable
reduction of the zero-energy single-particle spectral weight beginning
at temperatures as high as $T\lesssim 0.1t$, an effect that may be
detrimental to the existence of a ferromagnetic ground state at weak
coupling.
\end{abstract}

\pacs{71.10.Fd, 71.27.+a, 75.10.Lp}
\maketitle

\section{\label{sec1}Introduction}

Historically, the single-band Hubbard model was suggested
independently by Gutzwiller~\cite{gut}, Hubbard~\cite{hub} and
Kanamori~\cite{kan} to gain insight into the origin of metallic
ferromagnetism. However, despite enormous efforts~\cite{faz} that were
undertaken to find answers to this question, only a few reliable
results have been obtained even for this simplest possible microscopic
model. The Hubbard model also exhibits a variety of other competing
phases, including antiferromagnetic and superconducting phases.

The first exact results for ferromagnetism were obtained in the strong
coupling limit, $U\rightarrow \infty $\thinspace , by
Nagaoka~\cite{nag} and Thouless~\cite{tho} who showed that the ground
state of the Hubbard model with one hole or electron is ferromagnetic
at an infinitely large Coulomb repulsion. That result did not answer
the question of stability to a finite concentration of holes in the
thermodynamic limit. Improved bounds for the Nagaoka state have
recently been derived~\cite{hum-h} for various lattices in two and
three dimensions. Ferromagnetic ground states also occur if one of the
several bands of the model is dispersionless (so-called Lieb's
ferrimagnetism~\cite{lieb} and flat-band
ferromagnetism~\cite{mita}). Mielke and Tasaki proved the local
stability of ferromagnetic ground states in the Hubbard model with
nearly flat~\cite{tas} and partially filled~\cite{miel}
bands. Ref.~\cite{TasakiRev} contains a short review of these works as
well as new results for Hubbard models without the singularities
associated with flat bands. A review of results~\cite{voall} obtained
for the simple one-band Hubbard model in the last few years as well as
the results of Mielke and Tasaki suggest that the important
ingredients for ferromagnetism in that model are (a) an interaction
strength that is in the intermediate to strong coupling regime and (b)
a band that exhibits a strong asymmetry and a large density of states
near the Fermi energy or near one of the band edges.  Metallic
ferromagnetism at weak coupling, usually known as Stoner
ferromagnetism, has in fact been ruled out a long time ago by
Kanamori~\cite {kan} based on the argument that the renormalization of
the interaction strength brought about by $T-$matrix effects (Kanamori
screening) would never allow the Stoner criterion to be satisfied when
the density of states at the Fermi level $\rho \left( E_{F}\right)$ is
non singular. Physically, the largest possible effective interaction,
according to Kanamori, is equal to the kinetic energy cost for making
the two-particle wave function vanish when the two particles are at
the same site. That energy scales like the bandwidth $\rho \left(
E_{F}\right) ^{-1}$ so that the Stoner criterion $ 1-U\rho \left(
E_{F}\right) =0$ cannot be fulfilled. Quantum Monte Carlo calculations
confirm the quantitative nature of Kanamori's $T-$matrix
result~\cite{cdlt91}.

If there is Stoner-type ferromagnetism in weak to intermediate
coupling, it is thus clear that, as in the moderate to strong-coupling
case, one needs at least a singular density of states to overcome
Kanamori screening. An example of a model with singular density of
states at the Fermi energy as well as band asymmetry is the
two-dimensional (2D) Hubbard model with both nearest neighbor, $t$,
and next-nearest-neighbor, $t^{\prime }$, hoppings.  When the Fermi
energy is close to the Van Hove singularity the corresponding filling
is usually referred to as a \textquotedblleft Van Hove
filling\textquotedblright . At that filling, the Fermi surface passes
through the saddle points of the single-particle dispersion. There
are, however, other phases competing with ferromagnetism. At weak to
moderate values of the on-site Coulomb repulsion $U,$ for small
$t^{\prime }/t$ and close to half-filling, the 2D $t-t^{\prime }$
Hubbard model shows an antiferromagnetic instability. That instability
due to nesting is however destroyed~\cite{hovo} for a sufficiently
large ratio $t^{\prime }/t$ at weak interactions in two and three
dimensions, thus leaving room for other instabilities, including
$d$-wave superconductivity and metallic ferromagnetism.

The questions which we address in this paper are thus the
following. Can the asymmetry of the band and the large density of
states near the Fermi energy overcome the Kanamori argument and lead
to ferromagnetism in the 2D Hubbard model? What are the competing
phases? Most results on this problem (particularly for a square
lattice) fall into three different classes. (a) Momentum-cutoff
renormalization group (RG) methods~\cite{hame,hsfr}, and Quantum Monte
Carlo calculations~\cite{lihi} suggest that there is no evidence for
ferromagnetism. But the problem, in particular with numerical methods,
is that only very small system sizes can be used in a regime where the
size dependence is important. In addition, momentum-cutoff RG does not
allow the contribution of ferromagnetic fluctuations~\cite{hs}. So
these results should not be considered conclusive. (b) The second
class of results is based on Wegner's flow equations. They
show~\cite{hgw} a tendency towards weak ferromagnetism with $s^{\ast
}$-wave character (the order parameter changes sign close to the Fermi
energy). According to the flow equations calculations this phase
competes with other instabilities in the particle-hole channel, in
particular with the Pomeranchuk instability. The difficulty of those
weak-coupling calculations is that the $s^{\ast }$-magnetic phase
occurs at stronger coupling than the regime of validity of the second
order analysis in $U$ of the flow equations. (c) The third class
suggests clear evidence for ferromagnetic ground states. These works
include a projector Quantum Monte Carlo calculation with $20\times 20$
sites and the $T$-matrix technique~\cite{hsg}, a generalized random
phase approximation (RPA) including particle-particle
scattering~\cite{foh} and exact diagonalizations\ \cite{arr}. Similar
tendencies have been found by the authors of Refs.~\cite{aggv,ikk}
within the renormalization group and parquet approaches for the
so-called two-patch model. Honerkamp and Salmhofer recently
studied~\cite{hs} the stability of this ferromagnetic region at finite
temperatures by means of a Temperature Cutoff Renormalization Group
(TCRG)\ technique analogous to that used earlier for one-dimensional
systems~\cite{BC}. For $U=3$, they have found that the ferromagnetic
instability is the leading one for $\left\vert t^{\prime }\right\vert
>0.33\left\vert t\right\vert $ at Van Hove fillings with the critical
temperature strongly dependent on the value of $t^{\prime }$. When the
electron concentration deviates slightly away from the Van Hove
filling, the tendency towards ferromagnetism is cut off at low
temperatures and a triplet $p$-wave superconducting phase
dominates. The $U$-dependence of these ferromagnetic and
superconducting phases in the ground state has been studied in
Ref.~\cite{kaka} by means of the same TCRG at weak coupling.

In the present paper we study ferromagnetism and competing phases in
the $t-t^{\prime }$ Hubbard model at weak to intermediate coupling
by means of the two-particle self-consistent (TPSC)
approach~\cite{vt97}.  Antiferromagnetism and $d_{x^{2}-y^{2}}$-wave
superconductivity are the competing instabilities. The TPSC approach
is non-perturbative and applies up to intermediate coupling. It
enforces the Pauli principle, conservation laws and includes the
Kanamori screening effect. Comparisons with Quantum Monte Carlo
calculations have shown that TPSC is the analytical approach that
gives the most accurate results for the spin structure
factor~\cite{vct94}, the spin susceptibility~\cite{vt97} and the
$d_{x^{2}-y^{2}}$-wave susceptibility~\cite{klt02} in two
dimensions. Throughout the paper we consider the 2D $t-t^{\prime }$
Hubbard model at Van Hove fillings from weak to moderate couplings. We
determine the regions of the $T-t^{\prime }$ plane where the uniform
paramagnetic phase becomes unstable to various types of
fluctuations. We also estimate the electronic self-energy effects for
large $t^{\prime }$ where ferromagnetic effects are present. The next
section recalls the methodology. We then present the results and
conclude.

\section{\label{method}Two-particle self-consistent approach}

We consider the $t-t^{\prime }$ Hubbard model on a square lattice with
nearest ($t$) and next-nearest ($t^{\prime }$) neighbor hoppings 
\begin{eqnarray}
H &=&-t\sum_{\langle ij\rangle \sigma }(c_{i\sigma }^{\dagger }c_{j\sigma
}+h.c.)-t^{\prime }\sum_{\langle \langle ij\rangle \rangle \sigma
}(c_{i\sigma }^{\dagger }c_{j\sigma }+h.c.)  \notag \\
&&+U\sum_{i}n_{i\uparrow }n_{i\downarrow },
\end{eqnarray}
where $c_{i\sigma }^{\dagger }(c_{i\sigma })$ is the creation (annihilation)
operator for the electrons with spin projection $\sigma \in \{\uparrow
,\downarrow \}$, $U$ is the local Coulomb repulsion for two electrons of
opposite spins on the same site, and $n_{i\sigma }=c_{i\sigma }^{\dagger
}c_{i\sigma }$ is the occupation number. The bare single particle dispersion
has the form, in units where lattice spacing is unity, 
\begin{equation}
\varepsilon _{\mathbf{k}}=-2t(\cos k_{x}+\cos k_{y})-4t^{\prime }\cos
k_{x}\cos k_{y}.  \label{ek}
\end{equation}
This spectrum leads to a Van Hove singularity in the density of states
coming from saddle points of the dispersion relation that are located
at $\mathbf{k}=(0,\pm \pi )$ and $(\pm \pi ,0).$ The corresponding
energy is $\varepsilon _{VH}=4t^{\prime }$. In this paper we always
consider the case where the non-interacting chemical potential is
$4t^{\prime }$, so that the non-interacting Fermi surface crosses the
saddle points and the non-interacting density of states diverges
logarithmically at the Fermi energy. The filling corresponding to this
choice of chemical potential is a ``Van Hove filling''. For $t^{\prime
}=0$ and half-filling the Fermi surface is perfectly nested, namely
$\varepsilon _{\mathbf{k}+\mathbf{Q} }=-\varepsilon _{\mathbf{k}}$,
with $\mathbf{Q}=(\pi ,\pi )$, which leads to an antiferromagnetic
instability for $U>0$. The perfect nesting is removed for $t^{\prime
}/t\neq 0$. We work in units where Bolzmann's constant $k_{B}$ and
nearest-neighbor hopping $t$ are all unity.

The TPSC approach~\cite{vt97} can be summarized as
follows~\cite{avt03}. We use the functional method of
Schwinger-Martin-Kadanoff-Baym with source field $\phi $ to first
generate exact equations for the self-energy $\Sigma $ and response
(four-point) functions for spin and charge excitations (spin-spin and
density-density correlation functions). In such a scheme, spin and
charge dynamical susceptibilities can be obtained from the functional
derivatives of the source dependent propagator $G$ with respect to
$\phi $. Our non-perturbative approach then consists in two steps.

At the first level of approximation, we use the following two-particle
self-consistent scheme to determine the two-particle quantities: We
apply a Hartree-Fock type factorization of the four-point response
function that defines the product $\Sigma G$ but we also impose the
important additional constraint that the factorization is exact when
all space-time coordinates of the four-point function coincide. From
the corresponding self-energy, we obtain the local momentum- and
frequency-independent irreducible particle-hole vertex appropriate for
the spin response
\begin{equation}
U_{sp}={\frac{\delta \Sigma _{\uparrow }}{\delta G_{\downarrow
}}}-{\frac{\delta \Sigma _{\uparrow }}{\delta G_{\uparrow
}}}=U{\frac{\langle n_{\uparrow }n_{\downarrow }\rangle }{\langle
n_{\uparrow }\rangle \langle n_{\downarrow }\rangle }}.  \label{Usp}
\end{equation}
The renormalization of this vertex mainly comes~\cite{cdlt91,vt97} from
Kanamori screening~\cite{kan}. The double occupancy $\langle n_{\uparrow
}n_{\downarrow }\rangle $ entering this equation is then obtained
self-consistently using the fluctuation-dissipation theorem and the Pauli
principle. More specifically, the Pauli principle, $\langle n_{\sigma
}^{2}\rangle =\langle n_{\sigma }\rangle $, implies that 
\begin{equation*}
\langle (n_{\uparrow }-n_{\downarrow })^{2}\rangle =\langle n_{\uparrow
}\rangle +\langle n_{\downarrow }\rangle -2\langle n_{\uparrow
}n_{\downarrow }\rangle ,
\end{equation*}
while the fluctuation-dissipation theorem leads to an equality between the
equal-time equal-position correlation $\langle (n_{\uparrow }-n_{\downarrow
})^{2}\rangle $ and the corresponding susceptibility, namely 
\begin{equation}
\langle (n_{\uparrow }-n_{\downarrow })^{2}\rangle ={\frac{T}{N}}
\sum_{q}\chi _{sp}^{(1)}(q)=n-2\langle n_{\uparrow }n_{\downarrow }\rangle ,
\label{FD_Usp}
\end{equation}
where, using the short-hand $q\equiv (\mathbf{q},2i\pi mT)$, the summation
is over all wave vectors and all Matsubara frequencies with $T$ the
temperature, $n$ the electron filling, and $N$ the number of lattice sites.
The latter equation is a self-consistent equation for the double occupancy,
or equivalently for $U_{sp}$ in Eq.~(\ref{Usp}), because the
spin-susceptibility entering the above equation is 
\begin{equation}
\chi _{sp}^{(1)}(q)={\frac{\chi _{0}(q)}{1-{\frac{1}{2}}U_{sp}\chi _{0}(q)}},
\label{chi_sp}
\end{equation}%
where $\chi _{0}(q)$ is the particle-hole irreducible susceptibility
including the contribution from both spin components 
\begin{equation}
\chi _{0}(q)={\frac{2}{N}}\sum_{\mathbf{k}}{\frac{f(\varepsilon _{\mathbf{k}
})-f(\varepsilon _{\mathbf{k}+\mathbf{q}})}{2i\pi mT-\varepsilon _{\mathbf{k}
}+\varepsilon _{\mathbf{k}+\mathbf{q}}}},  \label{chi0}
\end{equation}%
with $f(\varepsilon )$ the Fermi-Dirac distribution
function. Eq.~(\ref{FD_Usp}) is also known as the local-moment sum
rule. The Green functions at this first level of approximation,
$G^{\left( 1\right) }$, contain a self-energy $\Sigma ^{\left(
1\right) }$ that depends on double-occupancy but since this
self-energy is momentum and frequency independent, it can be absorbed
in the definition of the chemical potential. In the above then, $
G^{\left( 1\right) }$ is the bare propagator and $\chi _{0}$ is the
bare particle-hole susceptibility both evaluated with the
non-interacting chemical potential $\mu _{0}$ corresponding to the
desired filling. The irreducible charge vertex $U_{ch}={\frac{\delta
\Sigma _{\uparrow }}{\delta G_{\downarrow }}}+{\frac{\delta \Sigma
_{\uparrow }}{\delta G_{\uparrow }}}$ strictly speaking is not
momentum and frequency-independent. Nevertheless, assuming for
simplicity that it is, it can be simply found by using the
fluctuation-dissipation theorem for charge fluctuations and the Pauli
principle,
\begin{equation*}
{\frac{T}{N}}\sum_{q}\chi _{ch}^{(1)}(q)=n+2\langle n_{\uparrow
}n_{\downarrow }\rangle -n^{2},
\end{equation*}%
with 
\begin{equation}
\chi _{ch}^{(1)}(q)=\frac{\chi _{0}(q)}{1+{\frac{1}{2}}U_{ch}\chi _{0}(q)}.
\label{chi_ch}
\end{equation}
The spin and charge susceptibilities obtained from Eqs.~(\ref{chi_sp})
and (\ref{chi_ch}) satisfy conservation laws~\cite{vt97,vct94}. This
approach, that satisfies the Pauli principle by construction, also
satisfies the Mermin-Wagner theorem: There is no finite-temperature
phase transition breaking a continuous symmetry. Nevertheless there is
a crossover temperature below which the magnetic correlation length
grows exponentially~\cite{vt97} until it reaches infinity at zero
temperature. Detailed comparisons of the charge and spin structure
factors, spin susceptibility and double occupancy obtained with the
TPSC scheme are in quantitative agreement with Quantum Monte Carlo
simulations for both the nearest-neighbor~\cite{vt97,vct94} and
next-nearest-neighbor~\cite{vdcvt95} Hubbard model in two dimensions.

In loop expansions, response functions are computed at the one-loop
level and self-energy effects appear only at the two loop
level. Similarly, in our case the second step of the approach gives a
better approximation for the self-energy. We start from exact
expressions for the self-energy with the fully reducible vertex
expanded in either the longitudinal or transverse channels. These
exact expressions are easy to obtain within the functional derivative
formalism. We insert in those expressions the TPSC results obtained at
the first step, namely $U_{sp}$ and $U_{ch}$, $\chi _{sp}^{\left(
1\right) }(q),\ \chi _{ch}^{\left( 1\right) }(q)$ and $G^{\left(
1\right) }(k+q)$ so that Green functions, susceptibilities and
irreducible vertices entering the self-energy expression are all at
the same level of approximation.  Then considering both longitudinal
and transverse channels, and imposing crossing symmetry of the fully
reducible vertex in the two particle-hole channels, the final
self-energy formula reads~\cite{malkpvt00,avt03}
\begin{eqnarray}
\Sigma _{\sigma }^{\left( 2\right) }(k)
&=&Un_{\bar{\sigma}}+{\frac{U}{8}}{ \frac{T}{N}}\sum_{q}\left[
3U_{sp}\chi _{sp}^{\left( 1\right) }(q)\right.  \notag \\ 
&&\left. +\
U_{ch}\chi _{ch}^{\left( 1\right) }(q)\right] G_{\sigma }^{\left(
1\right) }(k+q).  \label{selfen}
\end{eqnarray}
This self-energy~(\ref{selfen}) satisfies~\cite{vt97,malkpvt00,avt03}
the consistency condition between single- and two-particle properties,
$\mbox{Tr} (\Sigma ^{\left( 2\right) }G^{\left( 1\right) })=2U\langle
n_{\uparrow }n_{\downarrow }\rangle $. Internal consistency of the
approach may be checked by verifying by how much $\mbox{Tr}(\Sigma
^{\left( 2\right) }G^{\left( 2\right) })$ differs from $2U\langle
n_{\uparrow }n_{\downarrow }\rangle$. The results for single-particle
properties given by the self-energy formula~(\ref{selfen}) are in
quantitative agreement~\cite{vt97,malkpvt00,klpt03} with numerical
simulations at weak to moderate couplings. At temperatures much lower
than the crossover temperature where the correlation length increases
exponentially, the consistency condition signals that the method
becomes less accurate, although it does extrapolate in most cases to a
physically reasonable zero temperature limit~\cite{vt97}. In the
present paper, we will not present results below the crossover
temperature so we are always within the domain of validity. It should
be noted that the self-energy Eq.~(\ref{selfen}) takes into account
the fluctuations that are dominant already at the Hartree-Fock level,
namely the antiferromagnetic ones.

The above formalism can be extended~\cite{klt02} to compute pairing
correlations. Physically, the $d_{x^{2}-y^{2}}$-wave susceptibility
shows up after antiferromagnetic fluctuations have built up since it
is the latter that give some non-trivial momentum dependence to the
vertices. Momentum dependence of the vertices is absent in the bare
Hamiltonian and also at the first level of TPSC. It appears from the
momentum dependence of the self-energy at the second level of
approximation. In other words, our formalism physically reflects old
ideas about pairing by antiferromagnetic spin
waves~\cite{Scalapino95}. What it contains that is absent in other
formalisms is the possibility of suppression of superconductivity by
pseudogap effects also induced by antiferromagnetic
fluctuations~\cite{klt02}.

The mathematical procedure to obtain the $d_{x^{2}-y^{2}}$-wave
pairing susceptibility is as follows. Basically, the above steps are
repeated in the presence of an infinitesimal external pairing field
that is eventually set to zero at the end of the calculation. This
allows us to obtain the particle-particle irreducible vertex in Nambu
space from the functional derivative of the off-diagonal $\Sigma
^{\left( 2\right) }$ with respect to the off-diagonal Green
function. The $d$-wave susceptibility is defined by $\chi
_{d}=\int_{0}^{\beta }d\tau \left\langle T_{\tau }\Delta \left( \tau
\right) \Delta ^{\dagger }\right\rangle $ with the
$d_{x^{2}-y^{2}}$-wave order parameter $\Delta ^{\dagger
}=\sum_{i}\sum_{\gamma }g\left( \gamma \right) c_{i\uparrow }^{\dagger
}c_{i+\gamma \downarrow }^{\dagger },$ the sum over $\gamma $ being
over nearest-neighbors, with $g\left( \gamma \right) =\pm 1/2$
depending on whether $\gamma $ is a neighbor on the
$\widehat{\mathbf{x}}$ or on the $\widehat{\mathbf{y}}$ axis. $\beta
\equiv 1/T$, $T_{\tau }$ is the time-ordering operator, and $\tau $ is
imaginary time. The final expression for the $d_{x^{2}-y^{2}}$-wave
susceptibility is
\begin{widetext}
\begin{align}
\chi _{d}\left( \mathbf{q}=0,iq_{m}=0\right) & =\frac{T}{N}\sum_{k}\left(
g_{d}^{2}({\bf k}) G_{\uparrow }^{\left( 2\right) }\left( -k\right)
G_{\downarrow }^{\left( 2\right) }\left( k\right) \right) -\frac{U}{4}\left( 
\frac{T}{N}\right) ^{2}\sum_{k,k^{\prime }}g_{d}( {\bf k}) G_{\uparrow
}^{\left( 2\right) }\left( -k\right) G_{\downarrow }^{\left( 2\right)
}\left( k\right)  \notag \\
& \times \left( \frac{3}{1-\frac{U_{sp}}{2}\chi _{0}\left( k^{\prime
}-k\right) }+\frac{1}{1+\frac{U_{ch}}{2}\chi _{0}\left( k^{\prime }-k\right) 
}\right) G_{\uparrow }^{\left( 1\right) }\left( -k^{\prime }\right)
G_{\downarrow }^{\left( 1\right) }\left( k^{\prime }\right) g_{d}({\bf
k^{\prime }}),  \label{Suscep_d}
\end{align}
\end{widetext}
with $g_{d}(\mathbf{k})=\left( \cos k_{x}-\cos k_{y}\right) $ the form
factor appropriate for $d$-wave symmetry. The above expression
contains only the first two-terms of the infinite series corresponding
to the Bethe-Salpeter equation. It should be noted that the appearance
of $G^{\left( 2\right) }$ on the right-hand side of the equation for
the susceptibility Eq.~(\ref{Suscep_d}) allows pseudogap effects to
suppress superconductivity~\cite{klt02}. This effect is absent in
conventional treatments of pairing induced by antiferromagnons.

Since the crossover to the ferromagnetic ground state found in our
work appears at very low temperatures ($T\leq 1/200$), a large lattice
is required in order to avoid finite-size effects at those
temperatures. In the case of ferromagnetism, sensitivity of the
results to the lattice size at low $T$ can be avoided by making sure
that the lattice is large enough at any given temperature to reproduce
the weak $\ln T$ behavior of the bare particle-hole susceptibility
$\chi _{0}\left( \mathbf{q=}0,iq_m=0\right)$.  That singularity
reflects the singular density of states at the Van Hove filling. We
found that a $N=2048\times 2048$ lattice suffices to compute $\chi
_{0}$ entering the TPSC phase diagram. The sum over $\mathbf{q}$ in
Eq.~(\ref{FD_Usp}) can be performed on a coarser mesh without loss of
precision. To speed up the calculations and to overcome increasing
memory requirements, especially at low temperatures, we use the
renormalization group acceleration
scheme~\cite{Pao:1994}. Interpolation is used to obtain quantities at
temperatures that fall between those directly obtained with the
renormalization group acceleration scheme.

\section{\label{results}Weak ferromagnetism and other instabilities}

Without loss of generality, we can take $t>0$ and $t^{\prime }\leq
0$. In that case, the Van Hove filling is always at $n\leq 1$. The Van
Hove fillings $n\geq 1$ occur only when $t$ and $t^{\prime }$ have the
same sign, but this case can be mapped back to the situation $n\leq 1$
using the particle-hole transformation $c_{i\sigma }^{\dagger
}\rightarrow \left( -1\right) ^{i}d_{i\sigma }$ and $c_{i\sigma
}\rightarrow \left( -1\right) ^{i}d_{i\sigma }^{\dagger }$ where the
phase factor takes the value $+1$ on one of the two sublattices of the
bipartite lattice and $-1$ on the other sublattice. The sign of $t$
and $t^{\prime }$ can be changed simultaneously with the particle-hole
transformation defined by $c_{i\sigma }^{\dagger }\rightarrow
d_{i\sigma }$ and $c_{i\sigma }\rightarrow d_{i\sigma }^{\dagger
}$. Whenever a particle-hole transformation is performed, the
occupation number changes from $n$ to $2-n$. The Van Hove filling
vanishes at $\left\vert t^{\prime }\right\vert =0.5\left\vert
t\right\vert $ so we restrict ourselves to $\left\vert t^{\prime
}\right\vert <0.5\left\vert t\right\vert $. For larger $\left\vert
t^{\prime }\right\vert $ there is a change in Fermi surface topology.

We begin with the Random Phase Approximation (RPA) phase diagram in
the $T-t^{\prime }$ plane, then move to the TPSC crossover diagram and
conclude with a short section on effects that can be detrimental to
ferromagnetism.

\subsection{RPA phase diagram}

\begin{figure}[tbp]
\includegraphics{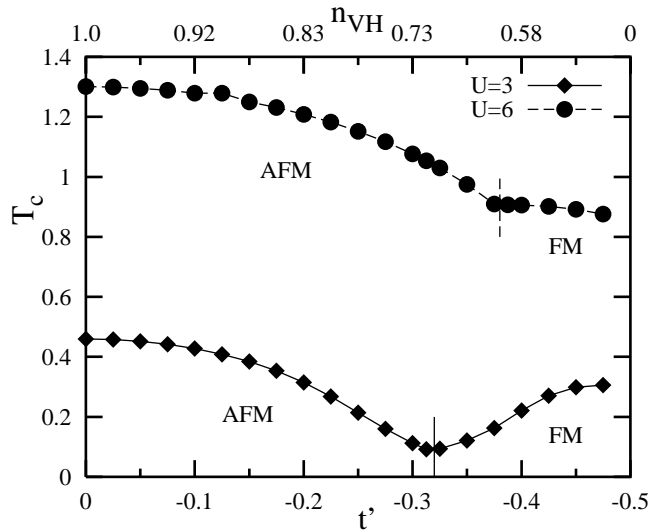}
\caption{The RPA critical temperature $T_{c}$ as a function of the Van
Hove fillings indicated on the upper horizontal scale and the
corresponding value of next-nearest-neighbor hopping $t^{\prime }$ on
the lower horizontal scale. The critical temperature $T_{c}$ is
determined from Eq.~(\protect\ref{criterion}). AFM stands for the
region where the uniform paramagnetic phase becomes unstable to
fluctuations at $\left( \protect\pi ,\protect\pi \right) $ while FM
is the region where the instability is at $\left( 0,0\right)
$. Vertical lines denote the boundary between AFM and ferromagnetic
phases. }
\label{Tc_rpa}
\end{figure}

Within RPA or mean-field, the transition temperature $T_{c}$ may be
found from
\begin{equation}
2-U\chi _{0}(\mathbf{q},0)=0,  \label{criterion}
\end{equation}%
where $\chi _{0}(\mathbf{q},0)$ is the zero-frequency limit of the
non-interacting particle-hole susceptibility given by
Eq.~(\ref{chi0}). In the case of ferromagnetism $\mathbf{q}=(0,0)$,
while $\mathbf{q}=\mathbf{Q}\equiv\left( \pi ,\pi \right) $ in the case of
commensurate antiferromagnetism. The temperature at which the uniform
paramagnetic phase becomes unstable to fluctuations at the
antiferromagnetic or at the ferromagnetic wave vector is plotted in
Fig.~\ref{Tc_rpa}. One should keep in mind that, in all cases, we are
speaking of spin-density waves, namely the local moment is in general
smaller than the full moment. Furthermore, for $|t^{\prime }|$
different from zero, the real wave vector where the instability occurs
is incommensurate. The question
of incommensurability is considered in more details in the TPSC
section. Note that in contrast to the case $U=3$, the ferromagnetic
critical temperature for $U=6$ does not increase with $t^{\prime }$,
it even decreases slightly. We do not explore the stability of the
various phases that could occur in mean-field theory below the
indicated transition lines.

\begin{figure}[tbp]
\includegraphics{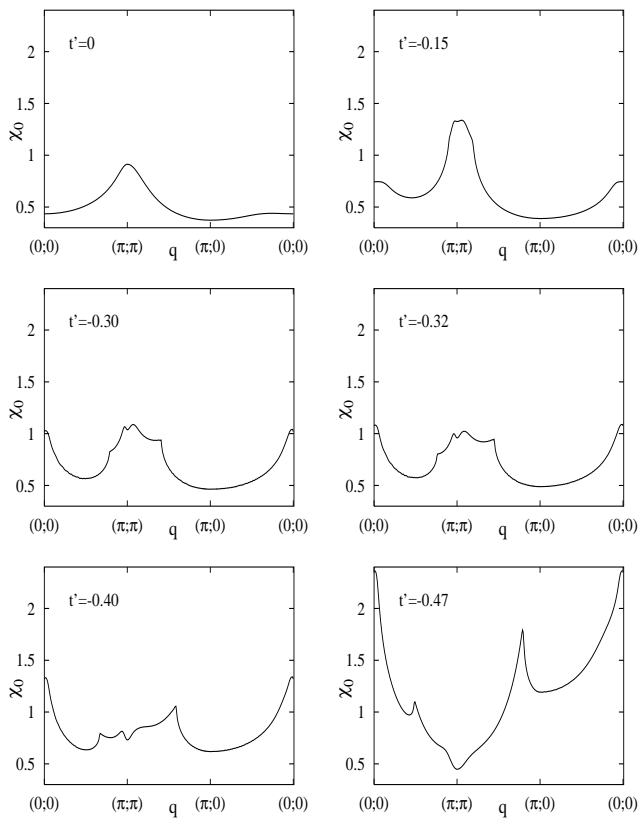}
\caption{The non-interacting particle-hole susceptibility
$\protect\chi _{0}$ at zero frequency as a function of wave vector
$\mathbf{q}$ along a path in the Brillouin zone is drawn for various
values of next-nearest-neighbor hopping $t^{\prime }$ at $T=0.01.$ The
filling is obtained by placing the chemical potential at the energy of
the Van Hove singularity for the given $t^{\prime }$.}
\label{chi0_q}
\end{figure}

In both RPA and TPSC, the wave vector where the instability first
develops is related to the $\mathbf{q}$-dependence of $\chi _{0}$. In
TPSC, it is not only the maximum value of $\chi _{0}\left(
\mathbf{q,}0\right) $ that determines the crossover temperature, but
the whole $\mathbf{q}$-dependence of $\chi _{0}$ that comes in the sum
rule Eq.~(\ref{FD_Usp}) for $U_{sp}$.  From the plot of $\chi _{0}$ as
a function of wave vector at $T=0.01$ in Fig.~\ref{chi0_q}, one can
see that at $t^{\prime }=0$ the antiferromagnetic wave-vector
$\mathbf{Q}$ leads to the largest value of $\chi _{0}$. With
increasing $|t^{\prime }|$ the maximum of $\chi _{0}$ is at an
incommensurate wave vector $\mathbf{Q}_{\delta}=(\pi -\delta ,\pi )$ close to
$\left( \pi ,\pi \right) $, while for large $|t^{\prime }|>0.32$
the maximum moves clearly to $\left( 0,0\right) $. For intermediate
negative values of the next-nearest-neighbor hopping $|t^{\prime
}|\sim 0.3$ the magnitudes of the susceptibility at $\left( 0,0\right)
$ and at $\left( \pi ,\pi \right) $ are comparable so the change in
the relative magnitude as a function of temperature is important.

The main deficiencies of RPA are (a) finite temperature phase
transitions in two dimensions that contradict the Mermin-Wagner
theorem, (b) an overestimation of the effect of $U$ on $T_{c}$ because
of the neglect of the renormalization of $U$ brought about by quantum
fluctuations (Kanamori screening). One can see from Fig.~\ref{Tc_rpa}
that the RPA critical temperature is quite a bit larger than the
crossover lines predicted by the TCRG (see Fig.~1 of
Ref.~\cite{hs}). The TPSC remedies these deficiencies.

\subsection{TPSC crossover diagram\label{TPSCcrossover}}

\begin{figure}[tbp]
\includegraphics{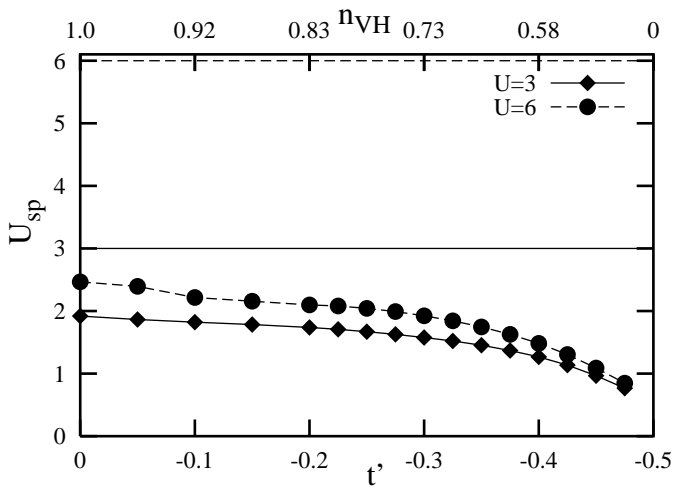}
\caption{Irreducible spin vertex $U_{sp}$ as a function of
next-nearest-neighbor hopping $t^{\prime }$ (or corresponding Van Hove
fillings on the upper horizontal scale) at $T=0.125$. Horizontal lines
at $U=3,\ 6$ denote the bare Hubbard repulsion $U$.}
\label{Usp_tp}
\end{figure}

We begin by considering the effective interaction $U_{sp}$ that plays
a crucial role in TPSC. In Fig.~\ref{Usp_tp} we plot $U_{sp}$ as a
function of $t^{\prime }$ as obtained from
Eqs.~(\ref{Usp}),~(\ref{FD_Usp}) and (\ref{chi_sp}). One can see that
Kanamori screening strongly renormalizes the effective
interactions. This weakly temperature dependent renormalization effect
is stronger for large $|t^{\prime }|$ in comparison with small
$|t^{\prime }|$. To explain this behavior we consider the sum rule
that determines $U_{sp}$, Eq.~(\ref{FD_Usp}). The main contribution to
the sum on the left-hand side of this equation comes from the small
denominator caused, for large $|t^{\prime }|$ by $\chi
_{0}(\mathbf{0,}0)$, and for small $|t^{\prime }|$ by $\chi
_{0}(\mathbf{Q},0)$. As the coefficient before the logarithm scales as
$[\sqrt{1-4(t^{\prime }/t)^{2}}]^{-1}$ for $\chi _{0}(%
\mathbf{0,}0)$, and as $\ln \left[ (1+\sqrt{1-4(t^{\prime
}/t)^{2}})/(2t^{\prime }/t)\right] $ for $\chi _{0}(\mathbf{Q},0)$, it
turns out that $\chi _{0}(\mathbf{0,}0)$ increases rapidly for
$|t^{\prime }|$ near $0.5$.  This means that $U_{sp}$ has to decrease
at large $|t^{\prime }|$ to satisfy the sum rule~(\ref{FD_Usp}) where,
in addition, the quantity $n-2\langle n_{\uparrow }n_{\downarrow
}\rangle $ on the right-hand side is a decreasing function of density
(and hence of $|t^{\prime }|$).

To find the crossover lines, we consider the zero-frequency limit of
the spin susceptibility given by Eq.~(\ref{chi_sp}) and the
$d_{x^{2}-y^{2}}$-wave pairing susceptibility given by
Eq.~(\ref{Suscep_d}) above. The crossover temperature $T_{X}$ for the
magnetic instabilities is chosen as the temperature where the
enhancement factor $\chi _{sp}(\mathbf{q},0)/\chi _{0}(\mathbf{q},0)$
is equal to $500$. We have checked that this corresponds to a magnetic
correlation length that fluctuates around $25$ lattice spacings for
$|t^{\prime }|$ between $|t^{\prime }|=0$ and $|t^{\prime }|=0.3
$. The crossover temperature $T_{X}$ is not very sensitive to the
choice of criterion because near and below the crossover region the
enhancement factor grows very rapidly (exponentially).

For pairing, we proceed as follows. Eq.~(\ref{Suscep_d}) contains only
the first two terms of the infinite Bethe-Salpeter series. The first
term (direct term) describes the propagation of dressed electrons that
do not interact with each other while the second term contains one
spin fluctuation (and charge fluctuation) exchange. This comes about
in our formalism because $\Sigma ^{\left( 2\right) }$ is a functional
of $G^{\left( 1\right) }$. We would have obtained an infinite number
of spin and charge fluctuation exchanges, in the usual Bethe-Salpeter
way, if we could have written $\Sigma ^{\left( 2\right) }$ as a
functional of $G^{\left( 2\right) }$. This is not possible within
TPSC. We have only the first two terms of the full series.  The
superconducting transition temperature in two dimensions is of the
Kosterlitz-Thouless type and is expected to occur somewhat below the
temperature determined from the Bethe-Salpeter equation (Thouless
criterion). We thus use, as a rough estimate for the transition
temperature for $d$-wave superconductivity, the temperature where the
contribution of the vertex part (exchange of one spin and charge
fluctuation) becomes equal to that of the direct part (first term) of
the $d$-wave pairing susceptibility~\cite{klt02}. In other words, we
look for the equality of the sign and the magnitude of the two terms
appearing in Eq.~(\ref{Suscep_d}).  This choice is motivated by the
statement that $1+x+\ldots $ resummed to $1/\left( 1-x\right) $
diverges when $x=1$.

\begin{figure}[tbp]
\includegraphics{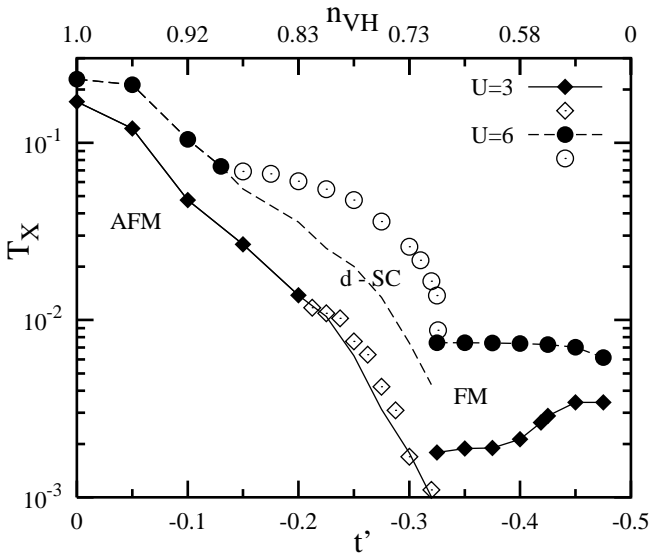}
\caption{The TPSC phase diagram as a function of next-nearest-neighbor
hopping $t^{\prime }$ (lower horizontal axis). The corresponding Van
Hove filling is indicated on the upper horizontal axis. Crossover
lines for magnetic instabilities near the antiferromagnetic and
ferromagnetic wave vectors are represented by filled symbols while
open symbols indicate instability towards $d_{x^{2}-y^{2}}$-wave
superconducting. The solid and dashed lines below the empty symbols
show, respectively for $U=3$ and $U=6$, where the antiferromagnetic
crossover temperature would have been in the absence of the
superconducting instability. The largest system size used for this
calculation is $2048\times 2048$.}
\label{Tc_TPSC}
\end{figure}

The TPSC phase diagram shows three distinct regions illustrated for
$U=3$ and for $U=6$ in Fig.~\ref{Tc_TPSC}: (a) for $t^{\prime }=0$,
the leading instability is at the antiferromagnetic wave vector and
for small non-vanishing $|t^{\prime }|$ it is at an incommensurate
wave vector close to $\left( \pi ,\pi \right) $. We will loosely refer
to that region as the region where antiferromagnetism dominates. (b)
For intermediate values of the next-nearest-neighbor hopping,
$d_{x^{2}-y^{2}}$-wave superconductivity dominates. (c) At large
negative $|t^{\prime }|>0.33$ a crossover to a magnetic instability at
the ferromagnetic wave vector occurs. Let us consider these different
regions in turn.

\begin{figure}[tbp]
\includegraphics{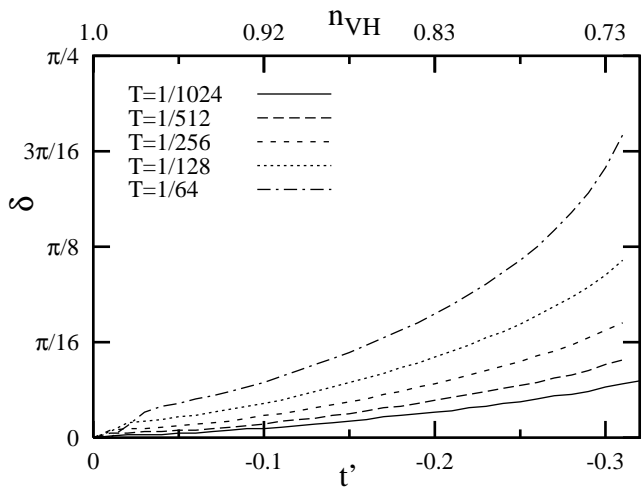}
\caption{Incommensurate wave vector $\mathbf{Q}_{\protect\delta
}=(\protect\pi -\protect\delta ,\protect\pi )$ where the maximum of
the non-interacting susceptibility is located as a function of
next-nearest-neighbor hopping $t^{\prime }$ at Van Hove
fillings. Different lines correspond to different temperatures. Given
$t^{\prime }$ and a crossover temperature in the TPSC phase
diagram, one can use the present figure to find out the
incommensurate wave-vector at which the instability first occurs.}
\label{delta}
\end{figure}

Near $t^{\prime }=0$, $T_{X}$ is relatively high and the
susceptibility near the antiferromagnetic wave vector grows most
rapidly. When we increase $|t^{\prime }|$, the crossover temperature
decreases because of reduced nesting of the Fermi surface. In TPSC the
wave vector of the instability is incommensurate for any finite value
of the next-nearest-neighbor hopping $|t^{\prime }|$ as can be
concluded from the structure of Eq.~(\ref{chi_sp}) and from the fact
that the non-interacting susceptibility with momenta
$\mathbf{Q}_{\delta }=(\pi -\delta ,\pi )$ is the largest when
$t^{\prime }\neq 0$. The incommensurate wave vectors are plotted in
Fig.~\ref{delta} as a function of $t^{\prime }$. One can see that the
degree of incommensurability is strongly temperature-dependent, and
that it increases with increasing temperature.

In the second region of the TPSC phase diagram $d_{x^{2}-y^{2}}$-wave
superconductivity is the leading instability. In this regime the
transition temperature to $d_{x^{2}-y^{2}}$-wave superconductivity
appears higher than the temperature at which the antiferromagnetic
correlation length becomes larger than about $25$. The latter
crossover lines are denoted by the solid $\left( U=3\right) $ and by
the dashed lines $\left( U=6\right) $ in Fig.~\ref{Tc_TPSC}. Note that
$d_{x^{2}-y^{2}}$-wave superconductivity is here induced by
incommensurate antiferromagnetic fluctuations. While high-temperature
superconductors are not generally close to Van-Hove singularities,
incommensurate dynamic spin fluctuations are concomitant with
$d_{x^{2}-y^{2}}$ superconductivity in these
compounds~\cite{Yamada98}.

Finally, the third regime occurs at $|t^{\prime }|>0.33$ where the
ferromagnetic susceptibility $\chi _{sp}(\mathbf{0},0)$ is the leading one
at low temperatures. Ferromagnetism occurs because of the diverging density
of states at the Van Hove singularity.

Note that for $U$ infinitesimally small the phase boundaries happen close
to zero temperature. Disregarding superconductivity for the moment,
let us consider where the phase boundary between antiferromagnetism
and ferromagnetism would be at small $U$. In that case, the asymptotic
behavior of the Lindhard function near $\mathbf{q=0}$ and
$\mathbf{q}=\mathbf{Q}$ is,
respectively,~\cite{hsg}
\begin{eqnarray*}
\chi _{0}(\mathbf{0},0) &\sim &\ln (1/\max (\mu ,T))/\sqrt{1-R^{2}} \\
\chi _{0}(\mathbf{Q},0) &\sim &\ln (1/\max (\mu ,T))\ln \left[
(1+\sqrt{1-R^{2}})/R\right],
\end{eqnarray*}
with $R\equiv 2t^{\prime }/t$ so that, looking at the equality of the
coefficients of the logarithms, one finds that the change from
antiferromagnetic to ferromagnetic behavior occurs at $|t^{\prime
}|=0.27$ instead of $|t^{\prime }|=0.33$ as found
above~\cite{NoteKatanin,aggv}. To understand the difference between
these two results, we need to look at subdominant corrections. For
example, a numerical fit reveals that $\chi _{0}(\mathbf{Q},0)\simeq
0.52+0.24\log _{10}(1/T)$. This means that for the leading term with a
logarithmic structure to be, say, about ten times larger than the
subdominant term, the temperature should be as low as $10^{-20}$.  The
corresponding $U$ (or $U_{sp}$) that satisfies $1=U($or
$U_{sp})\chi _{0}(\mathbf{Q},0)/2$ at this temperature is very small,
namely $0.4t$.  Therefore, unless $U$ is very small, the next leading
term plays an important role and a straightforward application of the
asymptotic form (taking only the leading term) is not justified. For
$U=6$ and $U=3$, for example, TPSC shows that near the
antiferromagnetic to ferromagnetic boundary, the crossover temperature
is of order $10^{-2}$ and $10^{-3}$ respectively. For this
temperature, the sub-leading term $0.52$ is comparable to the
logarithmic contribution.

The TPSC phase diagram is in qualitative agreement with the TCRG phase
diagram~\cite{hs}. In addition, the critical values $t_{c}^{\prime }$
for the stability of superconductivity and ferromagnetism are the same
in both approaches. But in contrast with the TCRG, ferromagnetism in
TPSC occurs at very low temperatures, and increasing $|t^{\prime }|$
does not lead to a dramatic increase in crossover temperature. One can
see from Fig.~\ref{Tc_TPSC} that the critical values of $t^{\prime }$
for the stability of ferromagnetism are unchanged for different $U$,
whereas the critical $\left\vert t_{c}^{\prime }\right\vert $ for the
stability of $d_{x^{2}-y^{2}} $-wave superconductivity decreases with
increasing coupling strength $U$.

The fact that the crossover temperature towards ferromagnetism depends
even more weakly on $t^{\prime }$ in TPSC than in RPA can be explained
by the following simple argument. Taking into account Kanamori's
improvement~\cite{kan} of the naive Stoner criterion for
ferromagnetism, we expect that the crossover temperature $T_{X}$ can be
roughly approximated by
\begin{equation}
T_{X}\sim T_{0}\exp {\left( -{\frac{1}{\rho (E_{F})U_{\mathrm{eff}}}}\right) 
},  \label{Tc_fm}
\end{equation}%
where $T_{0}$ is a constant, $\rho \left( E_{F}\right) =\chi _{0}\left( 
\mathbf{0},0\right) /2$ and $U_{\mathrm{eff}}$ is the renormalized effective
interaction ($U_{sp}$ in the case of TPSC). We have already explained in the
context of Fig.~\ref{Usp_tp} that the increase with $\left\vert t^{\prime
}\right\vert $ of the weight of the logarithmic singularity in the density
of states at the Fermi level leads to a decrease of $U_{sp}$, so the
crossover temperature is almost constant in TPSC. 

A distinctive feature of the TPSC phase diagram is that the crossover
to ferromagnetism generally occurs at much lower temperature than the
crossover to antiferromagnetism. This partially comes from the
peculiarity of the temperature dependence of the zero-frequency limit
of the non-interacting particle-hole susceptibility. To demonstrate
this, let us use, as an estimate for the crossover temperatures in
TPSC, the RPA criterion Eq.~(\ref{criterion}) with $U$ replaced by \
$U_{sp}$ and let us look for values of the temperature when the
left-hand side of that equation becomes small (it will vanish only at
zero temperature). At small $|t^{\prime }|$ the leading
non-interacting staggered susceptibility $\chi _{0}(\mathbf{Q},0)$
behaves like $(\ln {T})^{2}$ with temperature, while for $|t^{\prime
}|>0.33$ the leading non-interacting uniform susceptibility $\chi
_{0}(\mathbf{0},0)$ scales as $\left\vert \ln {T}\right\vert $. We
find that these susceptibilities have comparable size for temperatures
$T\gtrsim 1$, while the divergences of $\chi _{0}(\mathbf{Q},0)$ and
$\chi _{0}(\mathbf{0},0)$ begin respectively at $T<1$ and $T\ll
1$. Therefore, since the Stoner criterion Eq.~(\ref{criterion}) is
satisfied in RPA with bare $U=3,6$ at temperatures $T\gtrsim 1$, RPA
shows the same temperature scale for ferromagnetism and
antiferromagnetism. But in TPSC the strong renormalization of the
interaction strength $U_{sp}<U$ means that the crossover occurs for
larger values of $\chi _{0}(\mathbf{Q},0)$ and $\chi
_{0}(\mathbf{0},0)$, in a regime where they already have different
scales since $\chi _{0}(\mathbf{Q},0)$ for small $\left\vert t^{\prime
}\right\vert $ starts to grow logarithmically at much higher
temperature than $\chi _{0}(\mathbf{0},0)$ for large $\left\vert
t^{\prime }\right\vert $. Thus, the crossover to antiferromagnetism in
TPSC occurs at much higher temperatures than the crossover to
ferromagnetism.

Another interesting feature of the TPSC phase diagram at $U=3$ is that
the crossover temperature for antiferromagnetism is of the same order
of magnitude as that of the TCRG result of Ref.~\cite{hs}, whereas the
crossover to ferromagnetism is at much lower temperature than that
observed in the TCRG calculations. The naive explanation is as
follows. Let us assume that the approximate mean-field like expression
Eq.~(\ref{Tc_fm}) for $T_{X}$ has meaning both within TPSC and within
TCRG except that $U_{eff}$ has a different value in both
approaches. Simple algebra then shows that the relation between the
crossover temperatures for TPSC and TCRG in the ferromagnetically
fluctuating regime is
\begin{equation*}
{\frac{T_{\mathrm{FM}}^{\mathrm{TCRG}}}{T_{\mathrm{FM}}^{\mathrm{TPSC}}}}=\left(
{\frac{T_{0}}{T_{\mathrm{FM}}^{\mathrm{TPSC}}}}\right) ^{1-1/a},
\end{equation*}
with $a=U_{\mathrm{eff}}^{\mathrm{TCRG}}/U_{sp}$ characterizing the
different renormalizations of $U$ in both approaches. When $a=1$, both
crossover temperatures are equal. For $a>1$ the TCRG value for $T_{X}$
is larger than for TPSC while the reverse is true when $a<1$. Using
the numerical result~\cite{katan} for the TCRG effective interaction
at $U=3$ and $|t^{\prime }|\sim 0.45$ we have $a=1.4-1.8$. Then,
replacing $T_{0}$ by the bandwidth $8t$ and taking
$T_{\mathrm{FM}}^{\mathrm{TPSC}}=3.4\times 10^{-3}$ corresponding to
$|t^{\prime }|\geq 0.42$ we obtain the estimate
$T_{\mathrm{FM}}^{\mathrm{TCRG}}/T_{\mathrm{FM}}^{\mathrm{TPSC}}\approx
10-30$. This agrees with the crossover temperatures extracted from the
TPSC (Fig.~\ref{Tc_TPSC}) and the TCRG phase diagrams (Fig.~1 of
Ref.~\cite{hs}).  Similarly in the antiferromagnetically fluctuating
regime near $|t^{\prime }|=0$, we use the improved mean-field estimate
for $T_{X}$
\begin{equation*}
T_{X}\sim T_{0}\exp {\left( -\sqrt{8t/U_{\mathrm{eff}}}\right) },
\end{equation*}
to extract the following relation between the crossover temperatures 
\begin{equation*}
{\frac{T_{\mathrm{AFM}}^{\mathrm{TCRG}}}{T_{\mathrm{AFM}}^{\mathrm{TPSC}}}}=\left(
{\frac{T_{0}}{T_{\mathrm{AFM}}^{\mathrm{TPSC}}}}\right)
^{1-1/\sqrt{a}}.
\end{equation*}
Using the value~of $U_{sp}$ from the TPSC and the TCRG effective
interaction~\cite{katan} at $U=3$ and $|t^{\prime }|\sim 0.1$ we have
$a=1.0-1.4$. This leads to
$T_{\mathrm{AFM}}^{\mathrm{TCRG}}/T_{\mathrm{AFM}}^{\mathrm{TPSC}}\approx
1-2.5$ for $T_{\mathrm{AFM}}^{\mathrm{TPSC}}\sim 4\times 10^{-2}$ at
$|t^{\prime }|\sim 0.1$, which is in good agreement with the data
extracted from the phase diagrams.

As mentioned at the beginning of this subsection, the crossover
temperatures $T_{X}$ for the magnetic instabilities in TPSC have been
chosen such that the enhancement factor is equal to $500$. The
enhancement factor scales like the square of the correlation length
$\xi ^{2}$. For such large $\xi ^{2}$ the value of $T_{X}$ is rather
insensitive to the choice $500$ since the correlation length grows
exponentially. Our criterion for $T_{X}$ leads to a good estimate of
the real phase transition temperature with $\xi =\infty $ when a very
small coupling term is added in the third spatial direction. The
dependence of $T_{c}$ on coupling in the third dimension has been
studied, within TPSC, in Ref.~\cite{dvt96}. The latter reference also
contains expressions for the relation between the enhancement factor
and $\xi ^{2}$.  On the other hand, $T_{X}$ depends more strongly on
the precise criterion if we choose a moderate value of the enhancement
factor. In particular, the TPSC value of $T_{X}$ in the
antiferromagnetic fluctuation region increases by a factor two to five
if we choose $10$ for the enhancement factor, close to the
value~\cite{RGPD} chosen in Ref.~\cite{hs}. In this case, $T_{X}$
agrees essentially perfectly with the value obtained in the TCRG phase
diagram.

Note however that our estimate for the superconducting transition
temperature is smaller than that obtained with the TCRG of
Ref.~\cite{hs}.  Because in TPSC the pairing fluctuations do not feed
back in the antiferromagnetic fluctuations, this result suggests that
the feedback, usually included in TCRG, enhances superconductivity in
this region of the phase diagram. Such a positive feedback effect was
also found in the RG calculations of Refs.~\cite{bd01,hsfr}. On the other
hand, the RG approach of Ref.~\cite{aggv} suggests instead that
antiferromagnetism and superconductivity oppose each other. Some
particle-particle diagrams were however neglected in the latter
approach. In TPSC, antiferromagnetic fluctuations help
$d_{x^{2}-y^{2}}$-wave superconductivity as long as they are not
strong enough to create a pseudogap, in which case they are
detrimental to superconductivity~\cite{klt02}.

The above-mentioned renormalization group calculations were done in
the one-loop approximation without self-energy effects. By contrast,
in the RG work of Ref.~\cite{Zanchi}, self-energy effects showing up
at two loops were included in the calculation for the $t^{\prime }=0$
model. There, it was found that dressing the flow equations for AFM
and superconducting response functions with the one-particle wave
vector dependent weight factors $Z$ results in a reduction of both AFM
and superconducting correlations, the latter suppression being more
pronounced. Within TPSC, the momentum- and frequency-dependent
self-energy effects that appear in $G^{\left( 2\right) }$ in the
pairing susceptibility Eq.~(\ref{Suscep_d}) do tend to decrease the
tendency to pairing when AFM fluctuations become very strong at and
near half-filling~\cite{klt02}, in qualitative agreement with the RG
result~\cite{Zanchi}. In particular, in the presence of an AFM-induced
pseudogap, the tendency to superconductivity is decreased compared to
what it would be if we replaced $G^{\left( 2\right) }$ by $G^{\left(
1\right) }$ everywhere. (Such a replacement is not allowed within our
formalism). Because of the excellent agreement between TPSC at the
first level of approximation and Quantum Monte Carlo
calculations~\cite{vct94,vt97}, momentum and frequency dependent
self-energy effects are not expected to be very important for AFM
fluctuations unless we are deep in the pseudogap regime. They have not
been taken into account at this point. They might be more important in
the case of ferromagnetism, which is already a very weak effect in
TPSC. This is discussed in the following subsection.

\subsection{Additional effects that may be detrimental to ferromagnetism 
\label{DisfavorFM}}

The TCRG phase diagram~\cite{hs} is computed at the one-loop level.
Self-energy effects occur at the two-loop level. Similarly, self-energy
effects in TPSC are calculated at the second level of approximation. Since
analytical continuation of imaginary-time results is difficult at low
temperature, we estimate the quasiparticle weight with the help of the
quantity $z^{\prime }(T)$ defined in Refs.~\cite{vt96,vt97} by 
\begin{equation}
z^{\prime }(T)=-2G(\mathbf{k}_{F},\beta /2)=\int {\frac{d\omega }{2\pi
}}{\frac{A(\mathbf{k}_{F},\omega )}{\cosh (\beta \omega /2)}}.
\label{z}
\end{equation}
Physically, this quantity is an average of the single-particle
spectral weight $A(\mathbf{k}_{F},\omega )$ within $T$ of the Fermi
level ($\omega =0$). When quasiparticles exist, this is a good
estimate of the usual zero-temperature quasiparticle renormalization
factor $z\equiv (1-\partial \Sigma /\partial \omega )^{-1}$. However,
in contrast to the usual $z$, this quantity gives an estimate of the
spectral weight $A(\mathbf{k}_{F},\omega )$ around the Fermi level,
even if quasiparticles disappear and a pseudogap forms.
\begin{figure}[tbp]
\includegraphics{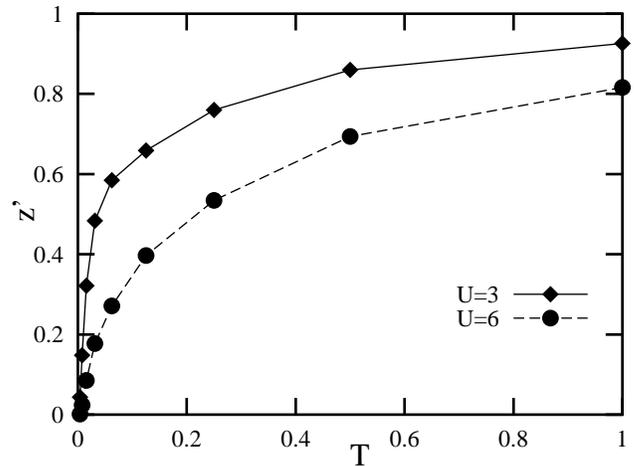}
\caption{Temperature dependence of $z^{\prime }(T)$ defined by Eq.\ (\protect
\ref{z}) at the Van Hove filling corresponding to $|t^{\prime }|=0.4$. }
\label{z_fig}
\end{figure}

Fig.~\ref{z_fig} shows the quasiparticle renormalization factor
$z^{\prime }$ at a value $|t^{\prime }|=0.4$ where ferromagnetic
fluctuations dominate at very low temperatures. One observes a
progressive decrease of $z^{\prime }\left( T\right) $ with decreasing
temperature. We checked that the single particle spectral function
$A(\mathbf{k}_{F},\omega )$ begins to show a small \textit{pseudogap}
at the temperature where $z^{\prime }$ begins to decrease
significantly. Since the ferromagnetic fluctuations are not yet strong
enough at that temperature to create a pseudogap, this effect is
completely driven by the singular density of states at the Van Hove
filling.  In other words, second-order perturbation theory should
suffice to observe the effect. The analogous feature was previously
found by one of the authors and his co-workers~\cite{Lemay:2003} in a
second-order perturbation study of the nearest-neighbor
two-dimensional Hubbard model at half-filling.  Self-energy effects
near Van Hove points have also been discussed in Ref.~\cite{ik02}. The
rather strong suppression of spectral weight at the Fermi wave vectors
for temperatures larger than the crossover temperature found in the
previous subsection would probably reduce the true $T_{X}$ or even
completely eliminate the possibility of a ferromagnetic ground state
if we could include the feedback of this self-energy effect into the
spin susceptibility.

The ferromagnetic fluctuation regime is also very sensitive to doping
within TPSC. In fact, deviations of the filling by $2-3\%$ away from
the Van Hove filling remove the crossover to the ferromagnetic regime.

There is also an argument that suggests that a Stoner-type
ferromagnetic ground state is unstable in the two-dimensional Hubbard
model. Within RPA in the ferromagnetic state~\cite{Doniach:1998}, the
spin stiffness constant for spin waves in the ferromagnetic state is
proportional to minus the second derivative of the density of states
at the Fermi level~\cite{n86}. Since the density of states as a
function of energy (away from the Van Hove filling) has a positive
curvature in two dimensions, that leads to a negative spin stiffness
constant and thus to an instability. This argument is based on the
non-interacting density of states. The pseudogap effect mentioned in
the previous paragraph changes the curvature of the density of states
at the Fermi level and may stabilize the ferromagnetic state.

\section{\label{concl}Conclusions}

As found within temperature-cutoff renormalization group
(TCRG)~\cite{hs,kaka}, TPSC suggests that ferromagnetism may appear in
the phase diagram of the 2D $t-t^{\prime }$ Hubbard model at Van Hove
fillings for weak to intermediate coupling. It is striking that the
overall phase diagrams of TCRG and TPSC have some close
similarities. As in TCRG, we find, Fig.~\ref{Tc_TPSC}, that for small
negative values of the next-nearest-neighbor hopping the leading
instability is a spin-density wave with slightly incommensurate
antiferromagnetic wave vector (Fig.~\ref{delta}). We could find
incommensurability at small $|t^{\prime }|$ only for very large
lattice sizes. The TCRG seems to indicate that very close to
$|t^{\prime }|=0$, the wave vector remains pinned at $\left( \pi ,\pi
\right) $~\cite{kaka} but that could be due to the fact that coupling
constants in TCRG represent a finite region in wave vector space and
hence very small incommensurabilities cannot be resolved. For
intermediate values of $|t^{\prime }|$ we also find
$d_{x^{2}-y^{2}}$-wave superconductivity. The precise value of
$|t^{\prime }|$ for the onset of $d_{x^{2}-y^{2}}$-wave
superconductivity depends somewhat on the criterion used for the
crossover temperature. One clear difference with TCRG, however, is
that the range of $|t^{\prime }|$ where superconductivity appears
increases with $U$ whereas it decreases with $U$ in
TCRG~\cite{kaka}. At large $|t^{\prime }|>0.33t,$ a crossover to
ferromagnetism occurs as a result of the diverging density of
states. TPSC cannot tell us what happens below the crossover
temperature, but that temperature is the relevant one in practice
since any small coupling in the perpendicular direction would lead to
a real phase transition.

The critical value for ferromagnetism, $|t^{\prime }|=0.33t$,
coincides with that found in TCRG~\cite{hs,kaka}. This value of
$|t^{\prime }|$ is smaller than that found within the $T-$matrix
approximation~\cite{hsg}, but that may be because of the cutoff to the
Van Hove singularity imposed by the small system sizes used in that
approach. The critical value for ferromagnetism, $|t^{\prime
}|=0.33t$, also differs from the value $|t^{\prime }|=0.27t$ obtained
in Ref.\ \cite{aggv} in the limit of zero temperature. We have
explained in Sec.\ \ref{TPSCcrossover} that for the crossover to occur
sufficiently close to $T=0$ for the arguments of Ref.\ \cite{aggv} to
be correct, one needs values of $U$ that are unrealistically small. At
finite $U $ (we studied $U=$ $3$ and $U=6$), subdominant corrections
to the logarithms shift the critical $|t^{\prime }/t|=0.27$ found by
Ref.\ \cite{aggv} to the value $|t^{\prime }/t|=0.33$ found by us and
TCRG.

The differences between TCRG and other approaches, as well as their
strengths and weaknesses, are well explained in
Refs.~\cite{hs,kaka}. The smaller temperature scale for crossover to
$d_{x^{2}-y^{2}}$-wave superconductivity in TPSC is a noteworthy
difference between our approach and TCRG~\cite{hs}. This may be due to
the fact that our calculations include self-energy effects which are
absent~\cite{Zanchi} in one-loop TCRG. But the most
striking difference is the temperature scale for ferromagnetism that
in our case remains extremely small away from the critical $|t^{\prime
}|=0.33t$.

We have shown that the low temperature scale for the crossover to
ferromagnetic fluctuations comes from Kanamori screening that strongly
renormalizes the effective interaction (this effect is smaller in the
antiferromagnetic regime). In TPSC this renormalization comes from the
constraint that the spin response function with $U_{sp}$ should
satisfy the local moment sum rule, Eq.~(\ref{FD_Usp}). This causes the
crossover temperature to ferromagnetic fluctuations to depend weakly
on $t^{\prime }$ and to remain small. As in the $T-$matrix
approximation~\cite{hsg}, Kanamori screening seems much stronger than
what is obtained with TCRG. The latter approach perhaps does not
include all the large wave vectors and large energies entering the
screening of the effective interaction.

Within TPSC then, the tendency to ferromagnetism seems very
fragile. In addition, we checked that in TPSC ferromagnetism
disappears for electron concentrations that are only very slightly
($2-3\%$) away from Van Hove fillings, in overall agreement with the
results of the TCRG~\cite{hs,kaka}.  So the question of the existence
of Stoner-type ferromagnetism at weak to intermediate coupling is not
completely settled yet, despite the positive signs and the concordance
of the most reliable approaches. We have estimated the electronic
self-energy effects for large $|t^{\prime }|$ and found that the
quasiparticle renormalization factor is reduced significantly at
temperatures $T<0.1$. As a result, the single-particle spectral
function $A(\mathbf{k}_{F},\omega )$ starts to show a small pseudogap
which, at high temperature, is completely driven by the singular
density of states, and not by the ferromagnetic fluctuations that
appear only at very low temperature.  This rather strong suppression
of spectral weight at the Fermi wave vectors for $T>T_{X}$ may further
reduce $T_{X}$ or even completely eliminate the crossover to a
ferromagnetic ground state. We have argued in Sec.~\ref{DisfavorFM}
that other factors could be detrimental to a ferromagnetic ground
state in two dimensions. In particular, as is the case with RG
calculations~\cite{hs,kaka}, a consistent treatment of the electronic
self-energy effects on the spin response function remains an open
issue.

Another interesting problem for future investigations is the question
of whether ferromagnetism could compete with the Pomeranchuk
instability, i.e.  a spontaneous deformation of the Fermi surface
reducing its symmetry from the tetragonal to the orthorhombic
one. Temperature cutoff RG~\cite{kaka,hsr02} disagrees with a
suggestion~\cite{hame,hgw,nm03} that this is one of the possible
leading instabilities of the 2D $t-t^{\prime }$ Hubbard model at Van
Hove fillings.

{\em Note added in proof:}
B.~Binz, D.~Baeriswyl and B.~Dou\c{c}ot [Ann. Phys. (Leipzig) 12,
(2003); cond-mat/0309645] have recently questioned the application of
one-loop renormalization group to ferromagnetism, suggesting that the
error produced by the one-loop approximation is of the same order as
the term which produces the ferromagnetic instability.

\begin{acknowledgments}
We are particularly indebted to A.~Katanin for sharing his unpublished
data with us and for numerous useful discussions. The authors also
thank M.~Salmhofer, C.~Honerkamp, C.~Bourbonnais, M.~Azzouz and
M. Gingras for valuable discussions. The present work was supported by
the Natural Sciences and Engineering Research Council (NSERC) of
Canada, the Fonds Qu\'{e}b\'{e}cois de la recherche sur la nature et
les technologies, the Canadian Foundation for Innovation and the Tier
I Canada Research Chair Program (A.-M.S.T.).
\end{acknowledgments}


\begin{thebibliography}{99}
\bibitem{RGPD} In the RG phase diagram of Ref.~\cite{hs} $T_{X}$ is defined
as the temperature where the coupling constant reaches $18t$, which
corresponds to the enhancement factor of approximately $20$ (private
communication with C.~Honerkamp).
\end{thebibliography}
\end{document}